\documentclass[aps,pra,nofootinbib,groupedaddress,twocolumn,final]{revtex4}

\def\sb{{\sf b}}

\def\cH{{\cal H}}

\newcommand\ket[1]{| #1 \rangle}

\newcommand\braket[2]{\langle #1|#2\rangle}

\newcommand{\beq}{\begin{equation}}
\newcommand{\eeq}{\end{equation}}
\newcommand{\rang}{\rangle}
\newcommand{\lang}{\langle}
\newcommand{\ovr}{\overline}

\begin{document}

\title{A simple unconditionally secure quantum bit commitment
  protocol via quantum teleportation\footnote{NOTE: This paper and the
  following two (quant-ph/0305143 and quant-ph/0305144) together provide a detailed description of various
  gaps in the QBC ``impossibility proof,'' as well as security
  proofs for four different protocols, QBC1, QBC2, QBC4, and QBC5. They also explain and correct some of the claims on my
  previous QBC protocols. This v3 on QBC5 is a minor improvement over v2.}}
\author{Horace P. Yuen}\email{yuen@ece.northwestern.edu}
\affiliation{Department of Electrical and Computer Engineering,
  Department of Physics and Astronomy, Northwestern University,
  Evanston, IL 60208}
\begin{abstract} By using local quantum teleportation of a
fixed state to one qubit of an entangled pair sent from the other
party, it is shown how one party can commit a bit with only
classical information as evidence that results in an
unconditionally secure protocol. The well-known ``impossibility
proof'' does not cover such protocols due to its different
commitment and opening prescriptions, which necessitate actual quantum
measurements among different possible systems that cannot be entangled
as a consequence.
\end{abstract}
\maketitle

It is nearly universally accepted that unconditionally secure
quantum bit commitment is impossible due to entanglement cheating.
For a brief summary of the problem and further references to the
literature, see Ref.~\cite{yue1}. In this paper, a concise
self-contained description of a protocol belonging to what we call
Type 5 QBC protocols, is presented together with a security proof.
Such protocols are based on two-way quantum and classical
communications within the framework of ordinary quantum mechanics,
without the use of additional constraints such as relativity or
super-selection rules, and in which only classical information is
committed as evidence.  The preliminary protocol QBC5p is a
two-stage protocol in which Babe sends Adam many maximally entangled
pairs of qubits $\cH_{\ell 1} \otimes \cH_{\ell 2}$, $\ell \in
\{1,\ldots,n\}$, on each of which a
unitary transformation $U_{\ell k}$ unknown to Adam has been applied to
$\cH_{\ell 1}$. Adam picks one pair $\cH_{\bar{\ell}1} \otimes
\cH_{\bar{\ell}2}$ and teleports to $\cH_{\bar{\ell}1}$ one of two known
orthogonal states in $\cH_3$ corresponding to the bit value
$\sb=0$ or 1, and commits the Bell-basis measurement result on
$\cH_{\bar{\ell}2} \otimes \cH_{\bar{\ell}3}$ to Babe. He opens by sending
$\cH_{\bar{\ell}1}$ and the remaining $n-1$ qubit pairs to Babe, who
verifies by making the corresponding projection measurements. This protocol
is $\epsilon$-concealing and can be extended to an
$\epsilon$--binding one, QBC5, in a sequence. The reasons underlying
the success of this protocol will be explained in this paper. But the {\em drastic difference}
between this protocol and the ones covered by the ``impossibility
proof'' (IP) should be evident. Aside from details, the main step of
QBC5 outlined above is quite simple, as is the idea behind it
which is the reverse of the usual formulation: measurement result
committed and quantum state given at opening instead of state
committed and measurement result given at opening. The teleportation
prevents Adam from entangling the opening possibilities.

Consider first the case of a single qubit pair. Starting from a fixed openly
known, say, ${\Psi}_{12}^- \equiv
\frac{1}{\sqrt{2}}(|\!\uparrow\rang _1 |\downarrow \rang _2 -
|\downarrow \rang _1 |\uparrow \rang _2 )$, Babe applies a unitary
$U_k$ known only to herself to $\cH_1$, and sends the
resulting $\cH_1 \otimes \cH_2$ in state $\Psi_k \equiv
(U_k \otimes I)\Psi_{12}^-$ to Adam. If Adam measures the Bell
basis
\begin{eqnarray*}
\Psi^{\pm} &\equiv& \frac{1}{\sqrt{2}} (|\uparrow\rang |\downarrow
\rang \pm |\downarrow \rang |\uparrow
\rang ),\\
\Phi^{\pm} &\equiv& \frac{1}{\sqrt{2}} (|\uparrow\rang |\uparrow
\rang \pm |\downarrow \rang |\downarrow \rang )
\end{eqnarray*}
on $\cH_2 \otimes \cH_3$ for any state $|\phi \rang$ on
$\cH_3$, he would obtain each result $i \in \{1,2,3,4\}$, with probability $\frac{1}{4}$ corresponding to $\Psi^{\pm}$ and
$\Phi^{\pm}$, with resulting state in $\cH_1$ given by:
\beq
U_k \sigma_i |\phi \rang
\label{eq:ustate}
\eeq where $\sigma_i$ are given by
$\sigma_1 = -\sigma_z, \sigma_2 = -\sigma_0 = -I, \sigma_3 =
-i\sigma_y, \sigma_4 = \sigma_x$ in terms of the Pauli spin
operators.  Let $\{|0\rang, |1\rang\}$ be an openly known
orthonormal basis of a qubit. To commit $\sb=0$ or 1, Adam may use
$\ket{\phi} = \ket{\sb}$ on $\cH_3$ and obtains
$U_k\sigma_i |\sb\rang$ on $\cH_1$ from (\ref{eq:ustate}), announcing
$i$ to Babe as the committed evidence. He opens by claiming b and
sending $\cH_1$ to Babe, who verifies by measuring the
one-dimensional projection of the state
$U_k\sigma_i|\sb\rang$. It is evident that the protocol is
perfectly concealing if she does not entangle the possible $U_k$:
with orthogonal states $|i\rang$ representing the classical
information committed by Adam, $\rho_0^B = \rho_1^B = |i\rang
\lang i |$.  Thus, her conditional probabilities for optimally
estimating the bit b on the basis of the committed evidence $i$
are $p(\sb|i) = \frac{1}{4}$ for both $\sb=0,1$.

For the corresponding binding proof of the situation in which Adam
opens one bit value perfectly, say opening $\sb=0$ with probability 1,
we will first show that Adam cannot cheat perfectly, i.e., his
optimal cheating probability $\ovr{P}^A_c$ of opening $\sb=1$ instead
is bounded away from 1.  This can be seen by exhausting all of his
possible actions.  First note that there is no entanglement
possibility for him either over the single state $|\sb\rang$,
or over the classical information $i$ even if it is represented by
$|i\rang$, as only a specific $i$ is accepted as legitimate.  In
order to cheat perfectly, Adam has several courses of action left.
The first one is for him to teleport $|0\rang$ to $U_k\sigma_i
|0\rang$ and then apply some $V^A$ to try changing it to
$U_k\sigma_i |1\rang$.  Another is for him to announce some $i$
without a Bell measurement and perform the teleportation on some
$|\phi \rang \in \cH_3$ with result $j$ when he opens b,
applying some $V^A$ to $\cH_1$ to obtain the proper opening
state.  The third is just sending some $|\phi\rang$ back to Babe
and try to open correctly. Finally, he may try to determine $U_k$. In the first case, his probability of
successful cheating is given by, for a given $i$,
\beq
P^A_c (1) =
\sum_k \lambda_k |\lang 1|\sigma_i^{\dagger} U^{\dagger}_kV^A U_k
\sigma_i |0\rang |^2
\label{eq:pac1}
\eeq where $\lambda_k$ is probability that
$U_k$ was employed by Babe. To obtain $P^A_c=1$, one would need
each of the terms in the sum of (\ref{eq:pac1}) to be one, which is impossible
if $\{ U_k\}$ is properly chosen, in particular if it is $\{ I, R_x, R_y, R_z \}$ where
$R_x$ is the rotation about the $\hat{x}$-axis of the qubit by an
angle $\pi/2$ \mbox{in its Bloch-sphere representation, etc.,}
\beq
\{U_k\} = \{I, R_x, R_y, R_z \}.
\label{eq:unitaries}
\eeq
 This is due to the fact that for each $U_k$, there are only one possible rotations
$V^A$ that makes $|\lang 1 |\sigma_i^{\dagger} U_k^{\dagger} V_B^A
U_k \sigma_i |0\rang |^2 = 1$ and there is no common rotation that
works for all $k$.  In the second situation the successful
cheating probability is, for a given pair $(i,j)$ and a given b,
\beq
P^A_c (2) = \sum_k \lambda_k |\lang b|\sigma_i^{\dagger}
U^{\dagger}_kV^A U_k \sigma_j |\phi \rang |^2.
\label{eq:pac2}
\eeq
Again,
equation (\ref{eq:pac2}) cannot be made equal to 1 with (\ref{eq:unitaries}) similar to the
case of (\ref{eq:pac1}). In the third case,
\beq
P^A_c (3) = \sum_k \lambda_k
|\lang b|\sigma_i^{\dagger} U^{\dagger}_kV^A  |\phi \rang |^2,
\label{eq:pac3}
\eeq
which clearly cannot be 1 as $V^A|\phi \rang$ is independent
of $k$. Note that the last two courses of action actually do not
allow him to open $\sb=0$ perfectly already. Also, other $\{ U_k \}$
can be used in lieu of (\ref{eq:unitaries}).

Note the role of the classical evidence $i$ and the role of
Babe's application of $U_k$ in preventing Adam's
perfect cheating.  Note also
that Adam has no entanglement cheating, in contrast to the IP formulation.  To summarize, Adam's optimum
cheating probability is some fixed number $\ovr{P}^A_c = p_A < 1$
for perfect \sb =0 opening. The exact value of $p_A$ depends on the
set $\{ U_k\}$ and its probability distribution, and can be
determined by solving the optimization problems corresponding to
his possible actions described above.  But there is no need to
determine this value for the purpose of proving the possibility of
obtaining unconditionally secure protocols.

Babe may entangle over the possible $U_k$ so that she keeps $\cH^C$ and sends $\cH_1 \otimes \cH_2$ for $|\Psi\rang
\in \cH^C \otimes \cH_1 \otimes \cH_2$ with
$|f_k\rang$ orthonormal in ${\cal H^{C}}$,
\beq
|{\Psi} \rang = \sum_k
\sqrt{\lambda_k} |f_k \rang (U_k \otimes I) | {\Psi}^-_{12} \rang.
\label{eq:psi}
\eeq
In this situation, one can cast the protocol in the IP formulation for
a fixed Bell-measurement result $i$ known to both parties by combining (\ref{eq:ustate}) and
(\ref{eq:psi}), and the protocol is not perfectly concealing. If Adam
commits only a partial Bell-measurement result, the protocol can still
be cast in the IP formulation if he only makes the partial (degenerate)
Bell-measurement, and not the full one, and then randomizes. To obtain
$\epsilon$-concealing, one can employ the following strategy --- Babe
sends Adam an ordered sequence of $n$ qubit pairs $\{ \cH_{\ell 1}
\otimes \cH_{\ell 2}\}$, each entangled in the form (\ref{eq:psi})
with $\lambda_{\ell k} = \frac{1}{4}$, $\ket{f_{\ell k}} \in
\cH^C_\ell$, and $\{ U_{\ell k}\}$ given by (\ref{eq:unitaries}). Adam randomly picks one of these
pairs, $\cH_{\bar{\ell}1} \otimes \cH_{\bar{\ell}2}$, and performs the
quantum teleportation. He commits the result $i$ of the
Bell-measurement as evidence, but not the name of the pair or
$\cH_{\ell 1}$ itself. He opens by sending in $\cH_{\bar{\ell}1}$ and all the other
$n-1$ qubit pairs. Babe verifies by measuring the projections onto
$U_{\bar{\ell}k}\sigma_i\ket{\sb}$ on $\cH_{\bar{\ell}1}$ (or the
projection onto (\ref{eq:psi}) in $\cH^C_\ell \otimes \cH_{\ell 1}$), and
$(U_{\ell k}\otimes I)\ket{\Psi^-_{\ell 12}}$ on the other pairs.

If Babe entangles pair by pair, it is easily seen that this preliminary protocol QBC5p is
$\epsilon$-concealing, as Babe has probability $\frac{1}{n}$ of guessing the
correct $\cH^C_{\bar{\ell}}$ for entanglement cheating to obtain a
cheating probability $\bar{P}^B_c > \frac{1}{2}$. If she does not
guess correctly, her cheating probability is $\frac{1}{2}$ as shown
above. Thus, her optimum $\bar{P}^B_c \le
\frac{1}{2}+\frac{1}{2n}$. If Babe entangles over all the pairs, a
calculation similar to that of Section IV in Ref.~\cite{yue3a} shows that,
conditioned on any of Babe's measurement results on her ancilla, the
same bound on $\bar{P}^B_c$ applies. This also follows from the fact
that pair-by-pair entanglement is entirely general in the present
case. See also Appendix A of this paper. At the same time,
Adam cannot cheat perfectly by operation on $\cH_{\bar{\ell}1}$ as
above, and he cannot entangle the different possibilities of
using any of the $n$ qubit pairs to obtain $i$ --- due to the
verification proecedure he could use only one qubit pair for
teleportation. One {\em cannot} entangle the different possibilities
involving measurements and no-measurements on different state
spaces. A classical measurement result is to be obtained, by Babe and
thus on Adam's qubits too, even if Adam does not perform the measurement
himself and just sends her a quantum state representing the result. The impossibility proof {\em assumes} that all opening
possibilities can be purified in an entanglement without a proof that
it is true in all possible situations. The above protocol provides an
example for which such an assumption is {\em not} valid.

Thus far the ``impossibility proof'' has already been
contradicted, since it asserts \cite{yue1} that whenever the protocol
is $\epsilon$-concealing, i.e., Babe's optimum cheating probability is
$\ovr{P}_c^B \sim \frac{1}{2}$, then $\ovr{P}_c^A \sim 1$ even in the
situation when he opens \sb =0 perfectly.  If the \sb =0 perfect opening
condition is relaxed, Adam has more possible actions open to him.
Exhaustively, they can be described by the following and their
combinations: approximate cloning of $\Psi_k$, using different
states than $|\sb\rang$ on $\cH_3$, announcing a
different $i$ from his Bell measurement, opening by sending a
different qubit from $\cH_1$ to Babe, and simply announcing a
different bit value at opening.  Two of these are already included
in the above analysis.  By continuity of all relevant functions on
a finite-dimensional space, it is readily seen that if Adam opens
on \sb =0 with probability $P_A(0)=1-\delta_1$, then he can cheat no
better than $\ovr{P}_c^A = p_A + \delta_2$ with
$\lim_{\delta_1\rightarrow 0}\, \delta_2=0$. While the optimum
${\ovr{P}_c^A}$ for arbitrary ${P_A(0)}$ remains to be determined,
there is no need to include it here as there does not seem to be
much interest in protocols where both ${P_A(0)}$ and ${P_A(1)}$
are not close to 1. In any event, the optimum tradeoff between
${P_A(0)}$ and ${P_A(1)}$ may be determined and the average
probability of opening correctly to Adam's choice can be brought
down to arbitrarily small level similar to the following
${P_A(0)=1}$ case.

Protocol QBC5' is obtained when the above QBC5p is repeated in a
sequence of $N$ such $n$-pairs, each consisting of the above protocol
step with the same $\sb$ opening. Adam's cheating probability
$\bar{P}^A_c = p^N_A$ can be made arbitrarily small for large
$N$. Babe's entanglement of the $Nn$ pairs clearly does not help her
cheat as she is left with the same small probability of matching the
correct $\cH^C_{\bar{\ell}}$ to the announced $i$.  However, her
$\bar{P}^B_c$ is improved by the $N$ repetitions, but it can be made arbitrarily small for any fixed $N$ by making
$n$ large. The quantitative treatment of a similar situation has been
given in Section VI of Ref.~\cite{yue5}. The key point is that Adam's
optimal cheating probability does not keep increasing with $N$, so
that arbitrarily large $N$ can be used to reduce $\bar{P}^B_c$ to any
level without corresponding increase in $\bar{P}^A_c$. Thus, the protocol QBC5 is $\epsilon$-concealing and $\epsilon$-binding for any desired
$\epsilon > 0$.

So far we have allowed Adam to cheat during both commitment and
opening but assumed that Babe is honest in sending Adam the ``legal'' states. Honesty during commitment in a
multi-stage protocol is an {\em assumption} of the ``impossibility
proof,'' which has to be relaxed for a truly secure protocol.  In
QBC5, Babe can cheat by, e.g., sending $|\psi_k\rang
|\psi_k\rang$, an unentangled state in $\cH_1 \otimes {\cal
H}_2$, so that the Bell measurement
probability on $|\psi_k\rang |\sb\rang$ in $\cH_2
\otimes \cH_3$ depends on b.  This kind of cheating, sending
in a different state other than one allowed in a multi-stage
protocol, can be handled in two different ways in general that
also apply effectively to the ``impossibility proof'' formulation
for certain protocols. The {\it first} way is to let one party
send in a large number of the allowed states so that the other
party can check for honesty on most of them and then use the rest.
In the present case, Babe would send Adam a large number $M$ of pairs, so Adam can set aside $m \ll
M$ of them, and asks Babe for the exact state in the other $M-m$
pairs. With or without her entanglement over such pairs, Babe has
to tell exactly what each state is, which Adam can
verify by a projection measurement. With $M$ sufficiently large,
it may be shown that the probability that any of the $m$ remaining
states is not a legal one allowed by the protocol, up to
entanglement by Babe, given all $M-m$ are, can be made arbitrarily
small for any $m$. The detailed analysis of this approach for the
present problem is given in Appendix A of this paper.

What if Babe is found cheating during such testing? Clearly, one
party can always refuse to cooperate in any protocol, which is
what repeated cheating and getting caught amounts to. It does not
alter the fact that the protocol allows arbitrarily small cheating
probability. One meaningful way to deal with repeated cheatings
applicable to realistic environments is to allow a fixed number
$n_c$ of cheating detection, beyond which the cheating party is
taken to be the loser in an essentially classical game-theoretic
formulation. The above $N$-ensemble can be generalized to deal
with such a formulation, but it is conceptually simpler to have
the following explicit game-theoretic formulation, our {\it
second} way to handle the use of non-allowed states during
commitment. In such a formulation, there is no $N$-ensemble sent
simultaneously.  Only one pair is sent, but the other party has a
large probability of choosing to check the validity of that state
instead of going further with the protocol. During the successive
trials, the other party may decide at any point to stop checking
and accept the state.  It is clear that with potentially an
unlimited number of trials, the probability of successfully
sneaking in a nonallowed state can be made arbitrarily small for
any fixed $n_c$.  With $n_c$ set to be zero, essentially no
cheating can be attempted at all without undue risk of losing the
game.  A quantitative description of this approach is given in
Appendix B of this paper. Note that neither of these approaches is
needed if one adopts the IP assumption that Babe is honest in
submitting only legal states of the protocol to Adam.

With either of these two ways, one can safely be assured that the
parties are honest during the exchanges of states or quantum
communications in the commitment phase of a multi-stage protocol,
as was assumed in the ``impossibility proof'' formulation.  Under
this condition, we have completed the concealing and binding proof
of the following simplification of the above protocol QBC5'.

\begin{center}
\vskip 0.1in
\framebox
{
\begin{minipage}{0.9\columnwidth}
\vskip 0.1in
\underline{PROTOCOL {\bf QBC5}}

{\small \begin{enumerate}
\item[(i)] Babe sends Adam $nN$ modified singlet pairs $(U_{\ell k}
\otimes I)\Psi^-_{\ell 12}$, $\Psi^-_{\ell 12} \in \cH_{\ell 1} \otimes
\cH_{\ell 2}$, $\ell \in \{1,\ldots,nN\}$, named
by their positions, with each $U_{\ell k}$, $k \in \{1,2,3,4\}$, randomly drawn from $\{U_{\ell k}\} = \{ I,R_x,R_y,R_z\}$, the $R$'s being $\pi/2$
rotations about the qubit axes.
\item[(ii)] Adam teleports the state $\ket{\sb}$ for the bit $\sb$ he
wants to commit to $n$ randomly selected pairs among the $nN$ ones,
committing to Babe the Bell-measurement results $i_m$, $m \in
\{1,\ldots,n\}$ without the corresponding names of the pairs.
\item[(iii)] Adam opens by sending $\{ \cH_{m1}\}$ and all the other
pairs together with their names to Babe, who verifies by corresponding measurements.
\end{enumerate}
\vskip 0.1in
}
\end{minipage}
}
\end{center}
\vskip 0.11in

To recapitulate the main reason for the security of QBC5: one cannot
purify the different possibilities of measuring different components
of a tensor product space while leaving the remaining part of the
space untouched. In QBC5, the tensor product space is $\bigotimes_\ell
(\cH_{\ell 1} \otimes \cH_{\ell 2})$ and the different components are
$\cH_{\ell 1} \otimes \cH_{\ell 2}$ indexed by $\ell$. Adam cannot entangle the
different possible actual measurements, and if he does not actually
measure, he could not open $\sb=0$ perfectly corresponding to the
committed $i$, which Babe could measure for him anyway in case he just
sends her the register containing the information $i$.

In addition to
teleportation, it is shown in Ref.~\cite{yue2} that the use of a
split entangled pair can yield an
unconditionally secure protocol. A general discussion on the scope
of the ``impossibility proof'' is given in Ref.~\cite{yue3}
together with the security proofs of several other protocols.

\appendix
\section{Honesty guarantee from an ensemble}
\par One general approach to guarantee with probability arbitrarily
close to 1 that a party $B$ is sending a ``legal'' state
from $\{\psi_k\}$ allowed by the protocol, or at least sending
the entangled superposition
\begin{equation}
\ket{\Psi}=\sum_k \lambda_k\ket{\psi_k}_B\ket{f_k}_C
\label{eq:super}
\end{equation}
for orthonormal $\ket{f_k}_C\in \cH^C$, is the following. She
sends in $N$ states, each randomly drawn independently from the
given allowed set $\{\psi_k\}$ and named, say, by its temporal
position. The other party $A$ randomly picks $N-n$ of such states
and asks $B$ to reveal them. After verifying that they are
correct, the probability that all $n$ remaining states are at
least of the form (\ref{eq:super}) can be made arbitrarily close to 1 by
proper choice of $n$, $N$ as follows.

Suppose $B$ mixes in states $\psi'$ that allows her to cheat
with probability $\bar{P}^B_c-\frac{1}{2}\geq\epsilon$
for a given $\epsilon$. Then the cheating detection probability
$\delta=1-|\braket{\psi'}{\psi_k}|^2>0$
minimized over the choice of $k$ and $\psi'$ is a fixed
number dependent only on $\epsilon$ and the protocol, independent
of $n$ and $N$. Suppose $B$ mixes in $m$ such $\psi'$ out
of the $N$ states she sends to $A$. We grant that $B$'s cheating
is successful if there is just one copy of $\psi'$ in the
$n$-group untested by $A$ and the measurements by him reveals no
different states from $\{\psi_k\}$. In order that there is a
non-vanishing probability found in the random $n$-group that $A$
sets aside, $m/N$ must be non-vanishing with $m/N\rightarrow{p}$
in the limit $N\rightarrow\infty$. Let $\alpha\equiv{n}/N$ so that
asymptotically for large $N, \alpha$ is the fraction of states
among the $N$ set aside. In order for $B$ to be able to cheat, one
has $m\alpha\geq 1$ for large $N$ because $m\alpha$ is the average
number of $\psi'$ in the $n$-group. The probability that
the cheating detection fails in the $N-n$ group is then
\begin{equation}
(1-\delta)^{m(1-\alpha)}\leq(1-\delta)^{\frac{1-\alpha}{\alpha}}\rightarrow
0,
\label{eq:testprob}
\end{equation}
which can be made arbitrarily small by
having $\alpha$ arbitrarily small that obtains with $N\rightarrow
\infty$ for fixed $n$. This argument can be completely quantified
via the hypergeometric distribution and the Chernov bound without
passing to the limit $N\rightarrow\infty$, although the limiting
argument suffices for the present purpose.

In QBC4, a simple modification is needed, in which Adam would need
Babe to send him $\cH^B_\beta$ together with the exact $\{ \ket{f_k} \}$ for checking that
the states are indeed $\ket{\Psi_\mu}\ket{\Psi_\nu}$ of the form
(\ref{eq:ustate}).

When the number of states $\{\ket{\psi_k}\}$ is equal to or larger
than the dimension of the state space $\cH^B$ sent to Adam, an
entanglement of the form (\ref{eq:super}) for each member of the
ensemble is entirely general. Any other entanglement can be obtained
by local transformation on $\cH^C$. This situation covers all our
applications.

The use of this method suffers from two disadvantages compared to the
next one in Appendix B via game-theoretic checking, but has the
advantage that no game payoffs need to be imposed. The first
disadvantage is that one needs to make sure that the evaluation of
$\bar{P}_A$ from Babe's ensemble takes into account the possibility,
depending on the $\ket{\psi_k}$'s, that Adam may be able to determine
it exactly with a nonzero probability. Of course, since he usually has
to return the unused one for Babe's verification, such a strategy may not improve his $\bar{P}^A_c$. In our qubit formulation such
possibility indeed does not arise.

The second disadvantage is that this approach gives Adam an ensemble
of choice for each individual $\ket{\psi_k}$ he is going to operate
upon, thus the opportunity of entanglement. Thus, one has to deal with
such entanglement possibility explicitly. This possibility does not
exist in QBC5, which is indeed the reason why it is secure as elaborated in this
paper. It does not exist in QBC4 because it is perfectly concealing,
but it does in our QBC1 and QBC2 which are dealt with in Ref.~\cite{yue2}.

\section{Honesty guarantee in a game-theoretic formulation}

During the commitment phase of a multi-stage QBC protocol
involving exchanges of quantum states, a party can try to cheat by
using ``illegal'' states not allowed in the protocol. In the
``impossibility proof'', this problem is not tackled by assuming
each party is honest, which is an unreasonable assumption in a
protocol that precisely does not place trust in either party. It
turns out, however, that a simple classical game formulation can
take care of this problem, up to entanglement of legal states, in
the following way.
\par\indent Suppose $B$ is giving $A$ a quantum
system in possible allowable states $\{ \psi_k \}$, to be further
processed by $A$. To make sure that the state is legal with a
probability arbitrarily close to 1, $A$ can ask $B$ to reveal the
state and check it by corresponding projection measurement. The
cheating detection probability is denoted by $p_d$, which is
determined by the illegal states $\psi '$ used by Babe that would
allow her to cheat beyond $\epsilon$, $\ovr{P}_c^B - \frac{1}{2}
\geq \epsilon$.  After checking, the whole protocol would need to
be repeated up to that point, but there is no question of resource
or efficiency in the present context.
\par Assuming first that the allowable number of cheats is $n_c=0$,
i.e., if $B$ is found cheating the game is over and $B$ will be
declared the loser. Such a situation occurs, e.g., when the
penalty of being found cheating is arbitrarily large.  We let $A$
employ a randomized strategy with a probability $p_a$ of accepting
$B$'s state without checking, independently from trial to trial
without loss of generality in this case. Thus $B$'s strategy can
also be represented by a probability $p_c$ of cheating at each
trial.  In an indefinitely long sequence of trials, $B$'s
successful cheating probability $P_C$ can be found as follows. The
probability $B$ will succeed at the first trial is $p_cp_a$, at
the second is $p_cp_a(1-p_a-p_cp_d)$, and at the nth is
$p_cp_a(1-p\ _a-p_cp_d)^{n-1}$. Thus, the total successful
cheating probability after $n$ trials is
\beq
P_C(n) = \frac{p_a}{p_a+p_d} [1-(1-p_cp_a-p_cp_d)^n].
\label{eq:ntrial}
\eeq
Similarly, the
probability that a legal state is accepted after $n$ trials is
\begin{eqnarray}
P_A(n) &=& \frac{p_a(1-p_c)}{p_a(1-p_c)+p_cp_d} \nonumber \\
&& \quad \times
  [1-(1-p_a+p_ap_c-p_cp_d)^n],
\label{eq:legalntrial}
\end{eqnarray}
and the probability a cheating would be detected is
\beq P_D(n) =
\frac{p_cp_d}{p_a+p_cp_d} [1-(1-p_a-p_cp_d)^n].
\label{eq:cheatdetect}
\eeq
Observing
that for fixed $p_d>0$, $P_D(n)\rightarrow 0$ is equivalent to
$p_c \rightarrow 0$ from (\ref{eq:cheatdetect}).  We have
\begin{eqnarray}
\lim_{\stackrel{P_D \rightarrow
0}{n \rightarrow \infty}} P_C(n) &=& 0,\\
\lim_{\stackrel{P_D \rightarrow
0}{n\rightarrow \infty}} P_A(n) &=& 1.\label{eq:lims}
\end{eqnarray}

With a large penalty for cheating,
  $P_D$ would be driven to zero from minimizing the average penalty.
Hence $p_c$ and $P_C$ are also driven to zero, as expected, with
$P_A \rightarrow 1$. Even if no penalty is imposed, $P_C(n)$ may
be made arbitrarily small for any $p_c > 0$ by making $n$ large
and $p_a/p_d$ small from (\ref{eq:ntrial}), while the corresponding $P_A(n)$
needs not be close to 1. Since a party has to cooperate and accept
a protocol if his/her security is guaranteed, as formalized by the
``Intent Principle'' of ref ~\cite{yue4} on protocol agreement, B
must pick a small $p_c$ to accept a viable protocol in this
situation.
\par With $n_c > 0$ but finite,
the same situation as above arises after $n_c - 1$ detections of
cheating, which would occur with probability arbitrarily close to
1 for a long enough number $n'$ of trials. It is also possible for
$A$ to terminate the game at any point by accepting the state
after a fixed number of trials $n_a$. Even though a more
complicated optimal strategy that is trial stage-dependent for
either party would then emerge, it would go over to the above for
large $n$.
\par If $\{ \psi_k \}$ is the set of allowable states,
this procedure only assures that the state $\psi \in \cH^{B}
\otimes \cH^C$ that $B$ sends in is of the form
\beq
|\psi
\rang = \sum_k \lambda_k |\psi_k {\rang}_{B} |f_k {\rang}_{C}
\label{eq:smallpsi}
\eeq
for orthonormal $|f_{k}\rang \in \cH^C$.  Often $B$ cannot
cheat
  already in such a case as the Type 5 ones in this paper, although in general
$B$ may still be able to cheat using (\ref{eq:smallpsi}). In some
  situations, as in QBC4 of Ref.~\cite{yue2}, one party may be asked
  to entangle as in (\ref{eq:smallpsi}) and later send in $\cH^C$ for
  the other party to verify the total entangled state.

In the full unconditionally secure protocol that requiring a
sequence, the above game can be repeated successively one by one.
The levels of each probability (\ref{eq:ntrial})-(\ref{eq:cheatdetect}) can be adjusted to
accommodate the desired level on the overall probabilities for the
$n$-sequence.

The generality of the entanglement (\ref{eq:smallpsi}) for our
protocols, as well as advantages and disadvantages of this
game-theoretic approach versus the ensemble approach, are briefly
discussed in the above Appendix A.

\begin{acknowledgments}This work was supported by the Defense
Advanced Research Projects Agency and the Army Research Office.
\end{acknowledgments}


\begin{thebibliography}{99}
\bibitem{yue1}H.P. Yuen, quant-ph/0210206.
\bibitem{yue3}H.P. Yuen, quant-ph/0305144.
\bibitem{yue3a}H.P. Yuen, quant-ph/0109055v1.
\bibitem{yue5}H.P. Yuen, quant-ph/0006109.
\bibitem{yue2}H.P. Yuen, quant-ph/0305143.
\bibitem{yue4}H.P. Yuen, quant-ph/0207089.

\end{thebibliography}
\end{document}